\documentclass[%
reprint,
 amsmath,amssymb,
 aps,
]{revtex4-2}

\usepackage{graphicx}

\begin{document}

\title{Synchronous Chip-to-Chip Communication with an Multi-Chip \\ Resonator Clock Distribution Network}

\author{Jonathan Egan, Max Nielsen, Joshua Strong, Vladimir Talanov,
  Ed Rudman, Brainton Song, Quentin Herr, and Anna Herr}

\affiliation{The authors are with Northrop Grumman Corp., Baltimore, MD 21240}

\thanks{This research is based upon work supported in part by the
  ODNI, IARPA, via ARO. The views and conclusions contained herein are
  those of the authors and should not be interpreted as necessarily
  representing the official policies or endorsements, either expressed
  or implied, of the ODNI, IARPA, or the U.S. Government.}

\date{28 April 2021}

\begin{abstract}
Superconducting digital circuits are a promising approach to build
integrated systems with high energy-efficiency and
computational density of the packaged chips.
In such systems, performance of the data link
between chips mounted on a multi-chip-module (MCM) is a critical
driver of performance. In this work we report a synchronous data link
using Reciprocal Quantum Logic (RQL) enabled by resonant clock
distribution on-chip and on the MCM carrier. The simple physical link
has only four Josephson junctions and 3\,fJ/bit dissipation, including
a 300\,W/W cooling overhead. The driver produces a signal with 35\,GHz
analog bandwidth and connects to a single-ended receiver via
20\,$\Omega$ Nb Passive Transmission Line (PTL). To validate this
link, we have designed, fabricated and tested two 32$\times$32\,mm$^2$
MCMs with eight 5$\times$5\,mm$^2$ chips connected serially and
powered with a traveling-wave clock, and with four 10$\times$10\,mm$^2$ chips
powered with a 2\,GHz resonant clock. The traveling-wave clock MCM validates
performance of the data link components, and achieved 5.4\,dB AC bias
margin with no degradation relative to individual chip test. The
resonator MCM validates synchronization between chips, with a measured
AC bias margin up to 4.8\,dB between two chips. The resonator MCM is
capable of powering circuits of 4 million Josephson junctions across
the four chips with a projected 10\,Gbps serial data rate.
\end{abstract}

\maketitle

Data movement efficiency drives overall power consumption and
performance in large-scale digital systems. Industry trends include
optimization of architectures and algorithms to minimize data
movement, and the use of energy-efficient optical interconnects on all
levels of the system from the chip, to the rack, to the data center
\cite{miller2017attojoule}. Superconducting digital technology has the
unique advantage of lossless, high bandwidth wires---essentially
optical properties---but using the native signal levels and the native
fabrication process
\cite{polonsky1993transmission}-\nocite{tanaka2005demonstration}\nocite{rafique2005optimization}\cite{shukla2019investigation}. This
avoids the overhead of low conversion efficiency associated with
optical. Superconducting on-chip interconnects can transfer signals
with 0.1\,fJ/bit total dissipation, including 300\,W/W cooling
overhead, over distances of 30\,mm \cite{talanov2021propagation}.

Superconducting lossless interconnect also enables integration of
chips and chiplets into a multi-chip module (MCM) with low overhead
and yield advantages \cite{herr2003ballistic}. This approach---long
recognized in the context of superconductivity---is now in full
alignment with the post-Moore industry trend towards new approaches to
mitigate unsustainable demand for computer power and unsustainable
hardware production cost. On the architecture side, the rise of
domain-specific hardware, heterogeneous systems, and chiplets integration
are moving the industry away from general purpose processors and
system-on-chip with ever-increasing integration density
\cite{shao2019simba}. On the hardware side, advanced packaging like
3D-IC enables new data protocols with increased pin and packaging
density \cite{farjadrad2019bunch}. All these advancements would be of
great effect when applied to superconducting technology due to
volumetric cooling. Cryogenic cooling is applied to the entire system
instead of to individual chips, enabling extremely high compute density
with all components of the superconducting system physically and
electrically close. An ExaFlop system might have the physical size of
three electronics racks \cite{holmes2013energy}.

The superconducting MCM is a primary component of a large scale
integrated cryogenic system. It consists of superconducting chips with
active logic and memory, mounted on a passive superconducting
carrier. Superconducting energy-efficient logic families such as RQL
\cite{herr2011ultra} and ERSFQ \cite{kirichenko2014ersfq} support
direct transfer of ps data pulses on passive transmission line
interconnects \cite{herr20138}, \cite{filippov2017experimental} (AQFP
technology requires costly energy conversion to RSFQ
\cite{china2016demonstration}). Flip-chip packaging with low
resistance bumps \cite{LincolnMCM}, \cite{kaplan2007high} means that
there is no energy penalty to transfer data off-chip. Total shoreline
bandwidth, which is a product of the number of signal lines and the
clock rate, is limited only by bump pitch.

Starting from the pioneering demonstration of 60\,Gbps chip-to-chip
communication \cite{herr2003ballistic}, there has been steady progress
in developing superconducting MCMs \cite{shukla2019investigation},
\cite{filippov2017experimental},
\cite{LincolnMCM}-\nocite{kaplan2007high}\nocite{hashimoto2005demonstration}\cite{narayana2012design}
and off-chip communication, including the recent demonstration of a
16-bit bus \cite{filippov2017experimental}. Typical demonstrations
involve a single chip mounted on the MCM with driver and receiver
located on the chip, and passive transmission lines on the
superconducting carrier. These demonstrations are important for
feasibility study of maximum data transmission
\cite{shukla2019investigation}, \cite{hashimoto2005demonstration} and
optimization of MCM link \cite{narayana2012design}. However,
single-chip circuits avoid the problem of timing synchronization that
is central to larger, real-world systems.

With typical clock frequencies for SFQ technologies of 2-60\,GHz
\cite{herr20138}, \cite{kashima202164}, \cite{ccelik2020fast} timing
synchronization between chips on MCM is an important aspect of the
system design. The local clocking in RSFQ logic families makes it
particularly challenging, and would appear to require complicated
clock recovery and channel de-skew. The AC-biased circuits of RQL have
the important advantage of enabling synchronous communication using a
resonant clock network on chip and between chips.

We report the first demonstration of a superconducting multi-chip MCM
with synchronous chip-to-chip links. The paper presents design and
test cycles 1) of nine 5$\times$5\,mm$^2$ chips on an MCM with a
traveling-wave
clock and characterized vs.\ clock rate, and 2) extension to four
10$\times$10\,mm$^2$ chips on an MCM with a global resonant clock to
demonstrate scalability.

\section{Chip-to-chip Synchronous Communication in RQL}

RQL technology is the first superconducting digital technology to
efficiently use AC power distribution using a resonant network. The
on-chip resonant clock and power network, implemented as a
metamaterial \cite{strong2022resonant}, is a proven enabler for
scaling to complex digital logic. The network has a Zeroth-Order
Resonance (ZOR) mode with infinite wavelength, providing uniform
amplitude and zero clock skew across the chip. Low skew enables
synchronous communication. For a multi-chip system the approach is
extended to a clock network across chips, with chip-to-chip
communication.


The MegaZOR is designed to the nominal resonance frequency of the
individual chips and equalizes the phases and amplitudes of the clock
signals across chips. The on-chip resonators can be considered small
perturbations to the MegaZOR. Analogous to the cavity perturbation
theory \cite{slater1946microwave}, \cite{pozar2011microwave6}, small
perturbations arising from materials or geometry shift the frequency
of the entire resonator at the expense of a slight change in the
partial eigenfields. The MegaZOR connects the chips with multiple
transmission line segments optimized for a uniform response in
amplitude and phase of the on-chip resonators. The entire structure
operates as a single resonator, even for spreads in the chip partial
resonant frequencies that may exceed the bandwidth of the resonance
peaks of the individual chips. The MegaZOR also transforms the small
22\,m$\Omega$ impedance of the on-chip resonators, four of which are
driven in parallel, to the 50\,$\Omega$ feedline impedance.

\begin{figure}
 \centering
 \includegraphics[width=3.4in]{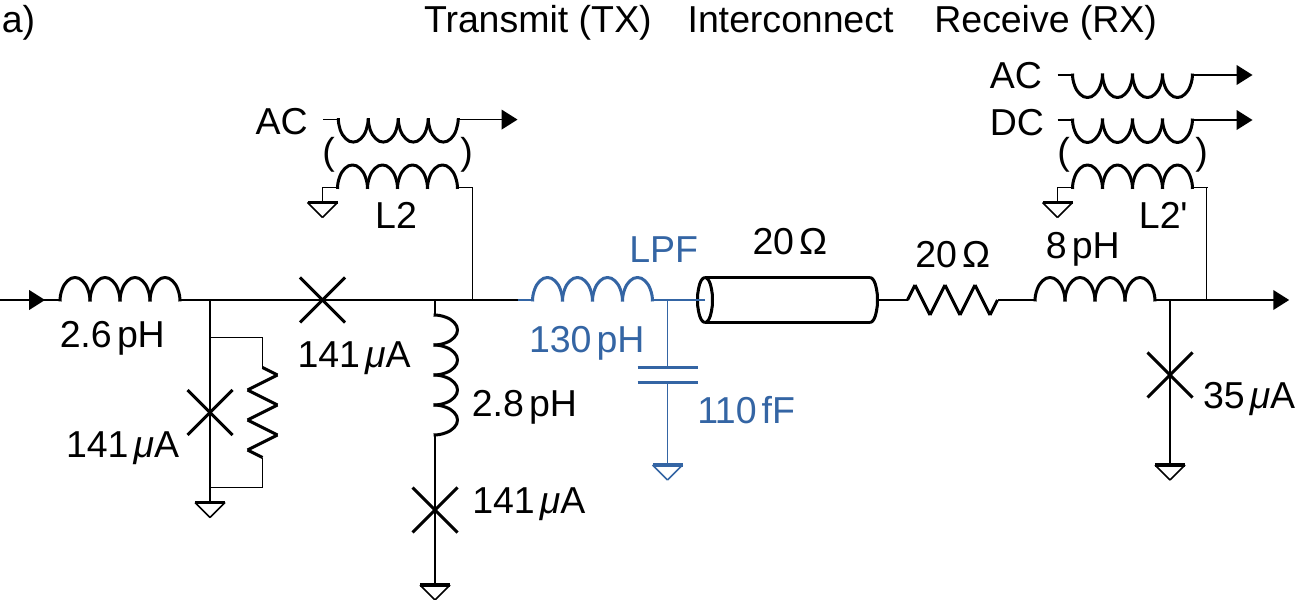} \\
 \includegraphics[width=3.5in]{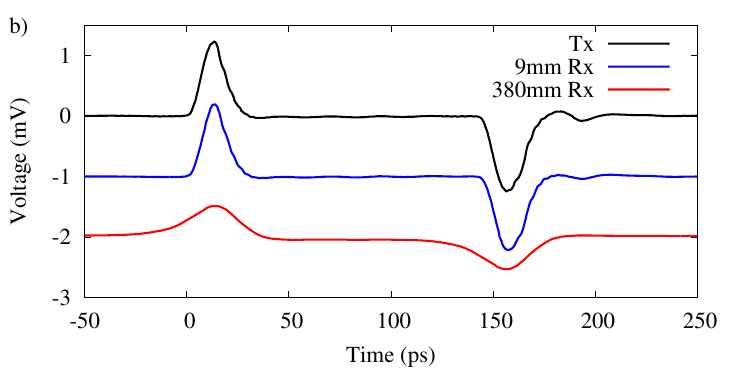} \\
 \caption{LBW synchronous communication link. a) The driver
   low-pass filters several underlying SFQ pulses generated by
   undamped, latching junctions to produce a Gaussian-pulse with 35\,GHz
   spectral bandwidth. The 20\,$\Omega$ interconnect is terminated at
   the receiver, which recovers the bipolar RQL-encoded data. b)
   Spectre-simulated waveforms at the driver output (Tx) and receiver input
   (Rx) show the attenuation and dispersion of the LBW pulse
   propagating over 9\,mm short and 380\,mm long PTLs. For clarity,
   propagation delay is subtracted from each waveform, and the
   waveforms are progressively offset by $-1$\,mV.
 \label{link}}
\end{figure}

The synchronous data link across the MCM requires that the chips have
the same frequency, phase, and amplitude within tight limits for
fabrication process variations. The delay between driver and receiver
is determined by signal propagation time on superconducting Passive
Transmission Lines (PTL) on the carrier. The clock phase in degrees of
the receiver, $\theta_r$, is
\[
\theta_r=\theta_d+\tau_{\mbox{\footnotesize PTL}}
\times f_{\mbox{\footnotesize res}} \times 360^{\circ},
\]
where $\theta_d$ is the phase of the driver,
$\tau_{\mbox{\footnotesize PTL}}$ is the electrical length of the PTL,
and $f_{\mbox{\footnotesize res}}$ is the frequency of the
clock. Latency may exceed one clock cycle, so the phase is wrapped,
e.g.\ if the PTL delay were 1025\,ps corresponding to a 100\,mm length
and the phase of the driver were $90^{\circ}$, then
$\theta_r=90+1025\,\mbox{ps} \times 5\,\mbox{GHz} \times
360^{\circ}=1935^{\circ}$, and subtracting $5 \times 360$ would yield
the nominal receiver phase of $\theta_r=135^{\circ}$. Synthesis and
timing analysis tools treat the MCM system as a single chip and assign
the phase of each receiver post-layout. Synchronous communication can
be used in systems where the fabrication-related skew between wires is
small relative to the timing window of the receiver, about one phase
of the RQL clock. For larger systems involving board-to-board
communication isochronous protocol is required, as described in
\cite{dai2021Isoch}.

An important component for long-range RQL data links is a driver with
improved bandwidth efficiency. Our Low Band-Width (LBW) driver has
about ten times less analog spectral bandwidth than the
single-flux-quantum (SFQ) pulse. Superconducting PTLs have some loss
and dispersion at RF frequencies which can degrade the signal over
distance. As first described in \cite{kautz1978picosecond} and
experimentally confirmed in \cite{talanov2021propagation}, a single
SFQ pulse with ps-scale width has a limited range of about 10\,mm on
typical Nb passive transmission lines of sub-micron width. Increasing
the pulse width increases the PTL range to meters, enabling
chip-to-chip and board-to-board interconnect.

Fig.~\ref{link}a shows the design of the LBW data link. All components
are AC-powered through inductive transformers. The LBW driver produces
7-10 SFQ pulses per bit, which requires the mutual inductance of the
L2 transformer to be about ten times larger than the standard for RQL
circuits. The LBW produces bipolar pulses 16\,ps wide, FWHM. The LBW
is triggered by a bipolar SFQ input and includes a floating buffer
junction to prevent back-action on the driving JTL. Circuit dynamics
are those of the self-resetting gate first described in
\cite{chan1975high}. Note that the width of the generated pulse is
still low compared to the clock period, giving the latching circuit
adequate time to reset and avoiding elevated bit-error-rate associated
with Josephson switching errors. The driver can
support up to 10\,Gbps throughput per link with 3\,fJ/bit dissipation,
including a 300\,W/W cooling overhead. The receiver achieves a wide
timing window of $80^{\circ}$ relative to AC clock, while maintaining
wide margins of $\pm 30\%$ on AC bias amplitude and about $\pm 50\%$
on individual junction critical current. Spectre simulations of the
LBW driver and transmission line, shown in Fig.~\ref{link}b, use a
dispersive propagation model for 1\,$\mu$m wide Nb 20\,$\Omega$ PTL on
SiO$_2$ that accurately models the high frequency components of the
pulse, as described in \cite{talanov2021propagation}.

Different fabrication processes were used for the chips for the two
circuit demonstrations. The 5$\times$5\,mm$^2$ chips were fabricated
in the ten-metal-layer SFQ5ee process developed by Lincoln Laboratory
\cite{tolpygo2019advanced}. The on-chip resonant clock-power network
occupies the bottom two metal layers and is separated from digital
logic by a superconducting ground plane. The
10$\times$10\,mm$^2$ chips were fabricated in the D-Wave six-metal
layer process \cite{berkley2010scalable} with a modified junction
critical current density of 10\,$\mu$A/$\mu$m$^2$. The process
features 0.25$\mu$m design rules for 5 bottom metal layers and
0.5$\mu$m design rules for the top layer. Here the on-chip resonant
clock-power network occupies the top two metal layers and is again
separated from the digital logic by a superconducting ground plane. In
both fabs, the chips are passivated with dielectric and use Ti/Au/Pt
pad metallization.

The MCMs for both demonstrations used a well-characterized process for
the carrier developed at Lincoln labs \cite{LincolnMCM}, and bump
bonds also developed at Lincoln. The SMCM4m four-Nb-metal MCM process
supports the design of 20\,$\Omega$ data PTLs and clock resonator
network, with good isolation provided by a superconducting ground
plane. The In bump-bond process supports up to 10,000 bumps with
15\,$\mu$m bump diameter, 35\,$\mu$m bump pitch, and 3\,$\mu$m
post-bond bump height.

\begin{figure}
 \centering
 \includegraphics[width=2.8in]{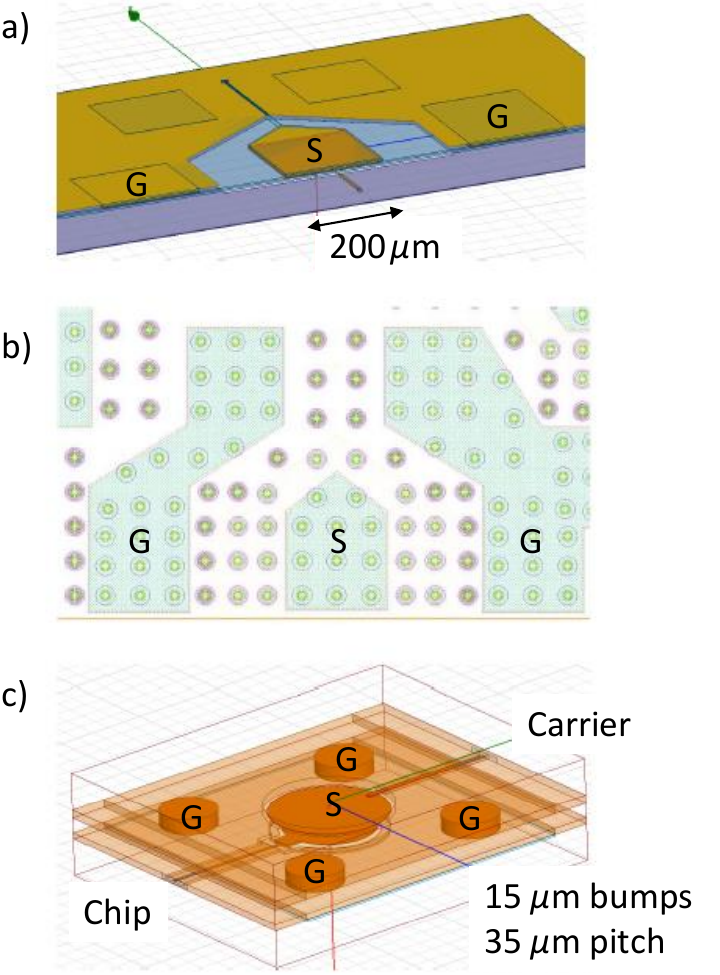}
 \caption{MCM interconnect. a) Pressure-contact probe packaging of
   individual chip used a standard pad template with large perimeter
   pads for clock and data. b) The same pads were used in the MCM
   attach for clock and I/O signals to room temperature. Multiple bumps
   were used for signal, ground, and mechanical support. c)
   High-bandwidth chip-to-chip interconnect used single, 15\,$\mu$m
   bumps for each signal and ground pad, on a 35\,$\mu$m pitch.
 \label{bumps}}
\end{figure}

Two kinds of pads were involved in attaching chips to the MCM, shown
in Fig.~\ref{bumps}. The high frequency, 20\,$\Omega$ chip-to-chip
interconnect used four ground pads surrounding each signal, and had a
single bump per pad. This transition was optimized using the HFSS 3D
field solver to achieve 350\,GHz analog bandwidth with $-20$\,dB
reflection. The chip-to-chip clock interconnect used large perimeter
pads with multiple bumps. Here the signal integrity requirements are
relaxed to a single tone bandwidth between 2-10\,GHz. The same large
perimeter pads are used to mount chips in the standard
pressure-contact dip probe and to connect MCM room temperature pads to
the chips. This configuration allows test of individual chips prior to
mounting on the MCM. In accordance with SMCM4m process design rules,
additional non-electrical bump bonds are added along the perimeter for
mechanical strength.

Process Control Monitor (PCM) and Time-of-Flight (TOF) resonators are
included on the MCM as circuit diagnostics for the MegaZOR and
interconnect PTL in order to track fabrication targeting
and spread. The PCM structures are characterized with four-port room
temperature resistance measurement. These low-cost tests are performed
for every MCM on every lot. These data indicate variations in the
metal width and thickness. Quarter and half wave TOF resonators
are tested at cryogenic temperatures using
a network analyzer to measure S-parameters. The frequency of the
resonance captures variations in the dielectrics and
metals. Correlating the PCM and TOF data gives the complete picture of
relevant fabrication process parameters.

\section{Multi-chip MCM with Traveling-Wave Clock}

\begin{figure}
  \centering
 \includegraphics[width=3.4in]{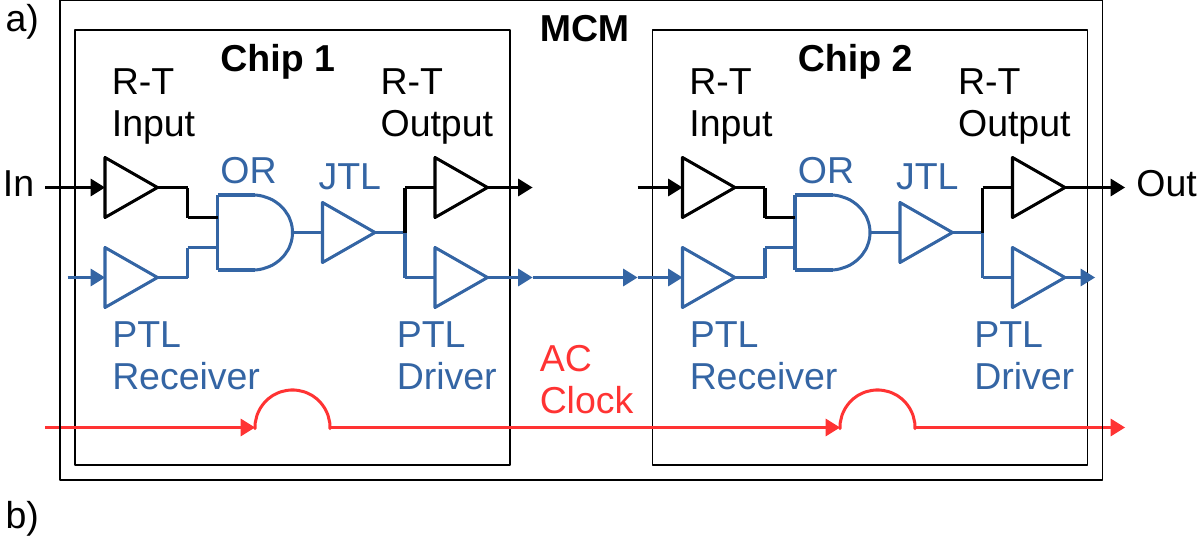}
 \includegraphics[width=2.7in]{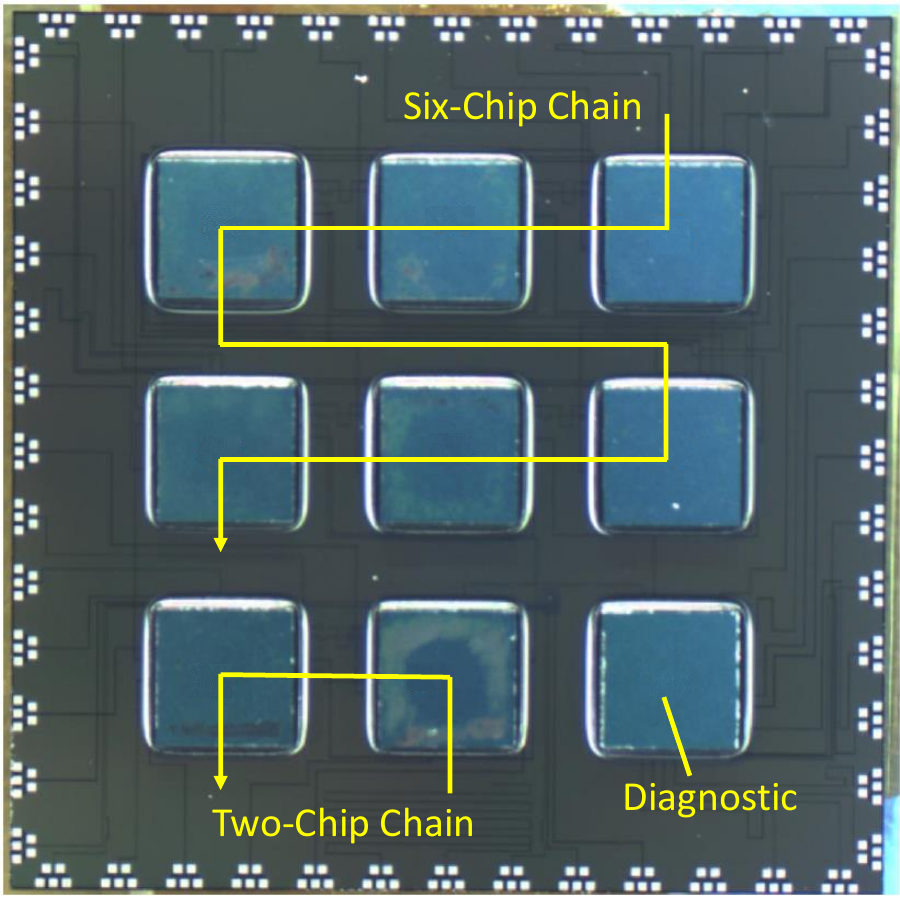}
 \caption{Synchronous data link with a traveling-wave clock. a) Schematic of
   two interconnected chips. Individual chips can be pre-tested using
   the room-temperature (R-T) interface prior to populating the MCM.
   The two-chip chain as shown uses the same R-T interface on the
   periphery, but uses the PTL interface chip-to-chip. Clock and data
   take similar paths on the MCM, with similar path lengths.  b)
   Microphotograph of the 32$\times$32\,mm$^2$ MCM populated with nine
   5$\times$5\,mm$^2$ chips, showing two-chip and six-chip data links.
 \label{the9}}
\end{figure}

The functionality of the on-chip circuitry, and the correctness of the
PTL model including transitions between chip and a carrier, was first
validated using a nine chip MCM circuit with a traveling-wave
clock. The traveling-wave clock has the advantage of tunable
frequency. The traveling-wave clock line and signal PTL delay are
designed for a 2\,GHz clock rate, but as the delays are the same the
design works across a large frequency range. The on-chip test circuit
has two data paths as shown in Fig.~\ref{the9}a, one connected to a
4\,mV output amplifier to pretest individual chips and the second
connected to a PTL driver for MCM tests.  On-chip circuitry designed
in this way also allows taps into input and output of individual chips
on MCM before testing the link through the multiple chips. In this
particular MCM design (see Fig.~\ref{the9}b) there two serial data
links going through 6 chips and 2 chips. The last chip was reserved
for a different experiment.

Individual chips and the MCM were tested in liquid Helium dip probes with
standard 32-pin and 64-pin pad templates. Tests were performed
at-speed with direct connection to room-temperature electronics, using
instrumentation and methods similar to that described in
\cite{egan2021true}.

\begin{figure}
 \centering
 \includegraphics[width=3.3in]{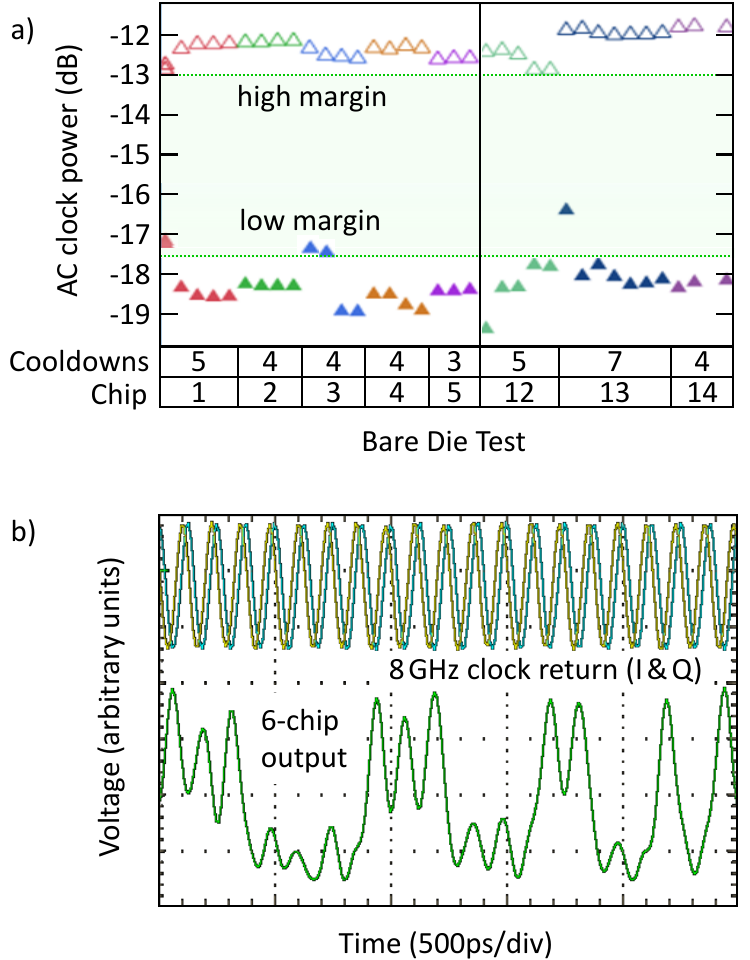}
 \caption{Traveling-wave-clock test results. a) Measured AC bias margins for
   the eight chips tested individually, at 2\,GHz and across multiple
   cooldowns. The data is grouped by a color per chip with multiple
   data points per cooldown. Typical margins are about 6\,dB. The
   shaded green region shows a common margin of about 4.5\,dB. b)
   Measured output waveforms through the six-chip chain on MCM at
   8\,GHz. The data pattern is a repetitive ``11110000111000110010''
   chirp, and the output is averaged.
 \label{fig4}}
\end{figure}

\begin{table}
\renewcommand{\arraystretch}{1.3}
\caption{Measured Margins for the Traveling-Wave-Clock MCM }
\label{9marg}
\centering
\begin{tabular}{cll}
  {\bf Build} & {\bf Frequency} & {\bf AC Margins} \\
\hline
    2 Chip Chain & \ \ \ 2\,GHz & \ \ \ 4.3\,dB \\
                 & \ \ \ 8\,GHz & \ \ \ 3.0\,dB \\
\hline 
    6 Chip Chain & \ \ \ 2\,GHz & \ \ \ 5.4\,dB \\
                 & \ \ \ 8\,GHz & \ \ \ 3.2\,dB \\
\hline
\end{tabular}
\end{table}

Fifteen individual chips were tested from the same wafer fabricated in the
SFQ5ee process, and eight were chosen to populate the chains on the
MCM. The individual-chip test for on-chip data paths showed up to
6\,dB ($\pm33\%$) of AC bias margin with common overlap between chips
of 4.5\,dB, as shown in Fig.~\ref{fig4}a. Operating margins across
multiple cooldowns showed about 1\,dB variation, which is within
the design specification for parasitic coupling to sequestered fluxes in
the moats. Similar measurements were taken on the two MCM links with
two and six-chip chains, at 2\,GHz and 8\,GHz clock rate.  Output and
clock-return waveforms at 8\,GHz are shown in Fig.~\ref{fig4}b. Both
links were functional with measured clock margins as entered in
Table~\ref{9marg}. The operating margins at 2\,GHz for both chains are
similar to those of the individual chips. Margins at 8\,GHz are
degraded by about 2\,dB, which indicates misalignment of the pulse
arrival time within the timing window of the receiver.

\section{Four-chip MCM with a Resonant Clock}

\begin{figure}
  \centering
  \includegraphics[width=2.7in]{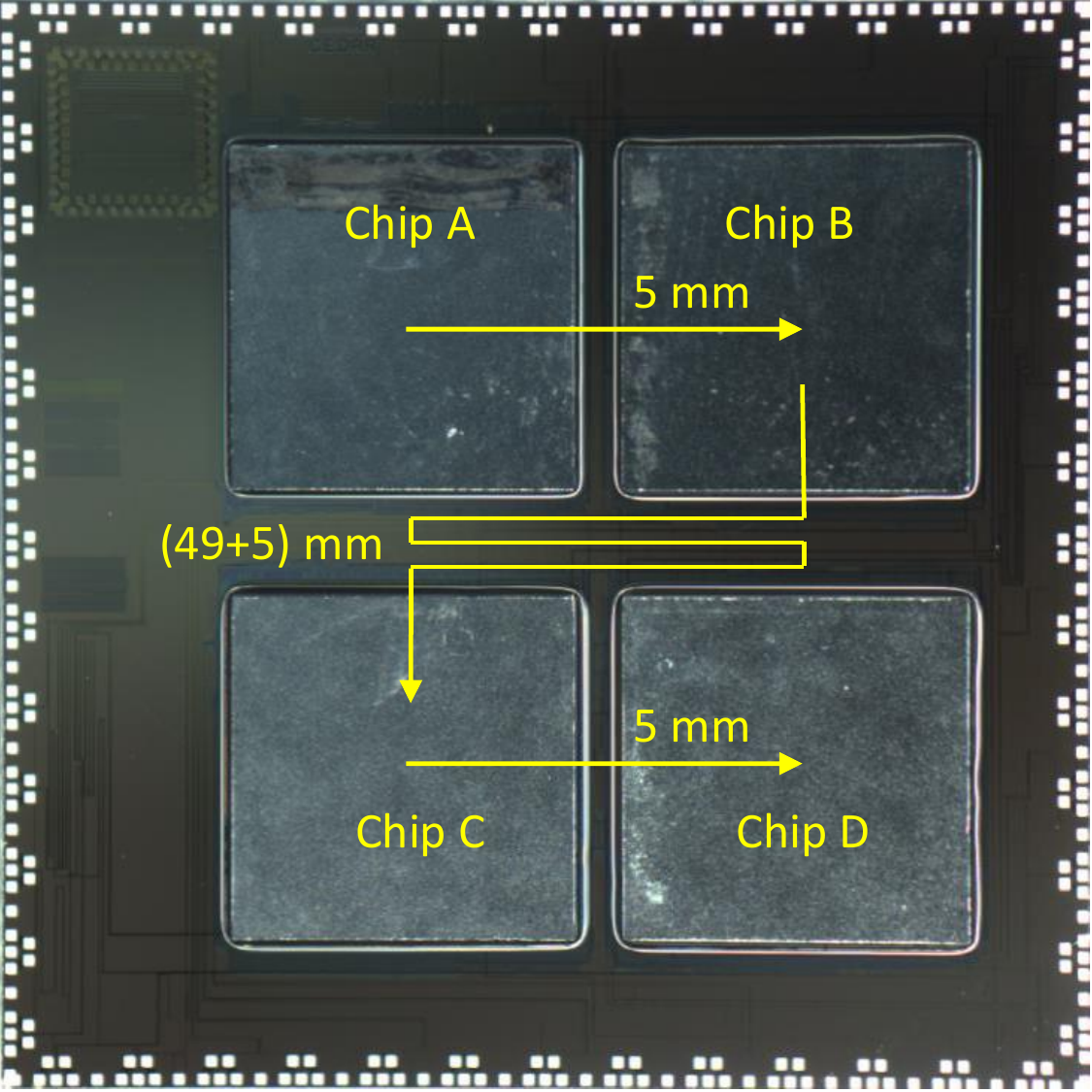}
 \caption{Microphotograph of the 32$\times$32\,mm$^2$ resonant-clock MCM
   populated with four 10$\times$10\,mm$^2$ chips, showing short and long
   signal paths between chips driven by a common resonator.
 \label{428photo}}
\end{figure}

The demonstration vehicle for synchronous communication using a
MegaZOR resonant clock is a 32$\times$32\,mm$^2$ MCM populated with
four 10$\times$10\,mm$^2$ chips, shown in Fig.~\ref{428photo}. This
build represents aggressive scaling due to the large on-chip ZOR clock
networks provisioning the entire active area of the
10$\times$10\,mm$^2$ chips. The whole system is capable of powering 4
million Josephson junctions. The increased scale stresses the
requirements for resonator amplitude and phase uniformity, bump
uniformity, and fabrication parameter spread. On-chip circuitry is
similar to that discussed in the previous section, but with an
additional synchronous data link providing a loop-back on-chip, in
parallel to the chip-to-chip data link. The four chips are connected
with two short 5\,mm PTLs and a long 54\,mm PTL that is one cycle
longer at 2\,GHz. Allocation of extra taps along the data path enables
test of all four chips individually, in pairs and in threes including
chip-to-chip communication, and complete test of data propagation
through all four chips.

\begin{figure*}
 \centering
 \includegraphics[width=6.4in]{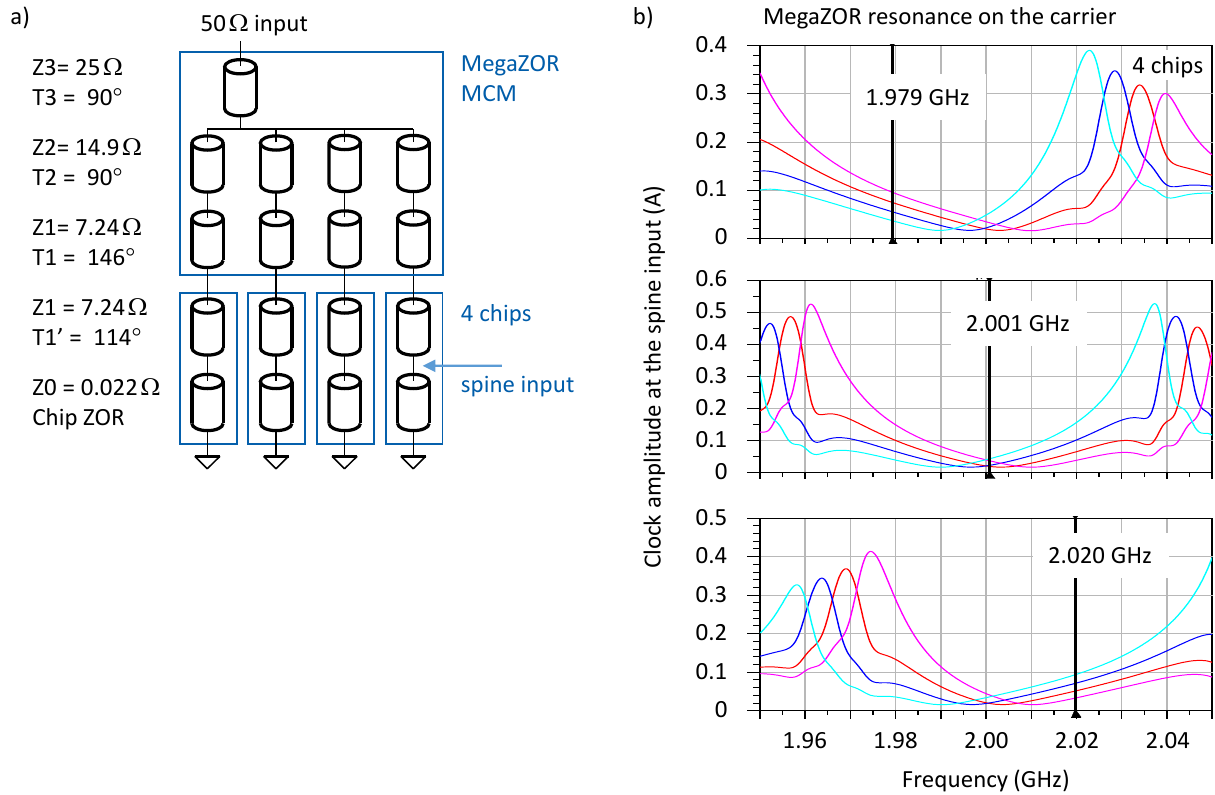}
 \caption{MegaZOR design. a) Schematic showing four identical branches
   of transmission line segments, each connected to the ZOR resonator
   on an individual chip. The segment length is given in degrees at
   the resonance frequency. b) Simulated current amplitude at the
   spines of on-chip resonator as function of frequency given for
   three different cases. The middle plot corresponds to the case
   with total $1\%$ mismatch between individual chip resonances
   frequencies with 2 chips being at $\pm 0.5\%$ and 2 chips being at
   $\pm 0.167\%$ from central frequency of 2\,GHz. Maximum spine
   current amplitude at the drive frequency is 42\,mA. The top and
   bottom plots correspond to the cases with added carrier MegaZOR
   frequency variation of $\pm 1.5\%$ on top of the frequency
   distribution between chips. Maximum spine current amplitude at the
   drive frequency is 50\,mA.
 \label{fig6}}
\end{figure*}

The schematic of the MegaZOR on the MCM is shown in
Fig.~\ref{fig6}a. It consists of four branches each connected to the
individual chip resonator. Each branch consists of two segments with
$90^{\circ}$ and $260^{\circ}$ transmission line 
resonators, which positions the connections between circuit segments
at voltage anti-nodes. The impedances of the segments are adjusted to
achieve a total loaded quality factor of resonator of
$Q_{\mbox{\footnotesize load}} = 73$. This represents a nice
compromise between minimizing input power and maximizing the
bandwidth.

There are two distinct MegaZOR networks for I and Q, electrically
identical but with different physical layout. The I/Q clock networks
are used to make the RQL four-phase
clock. In the current
MCM process with four metal layers, the resonant network occupies the first
metal layer and is separated from data transmission lines by a
superconducting ground plane in the second metal layer. This
constrains the layout to minimize the number of crossovers between the
I and Q networks. Physical layout components were simulated using
Ansys HFSS to determine impedances and propagation speeds, and
sensitivities to process variations. The top-level netlist was
simulated using Keysight's Advanced Design System (ADS).

The on-chip resonant clock network is described in \cite{strong2022resonant}.
Fig.~\ref{fig6}b shows simulated current amplitudes at the spines of
the four chips in the worst-case assumption of chip frequency mismatch
range of $\pm 0.5\%$ and three different carrier frequencies at
nominal, 2\,GHz, and $\pm 1.5\%$ from nominal. In the ideal case when
there is no mismatch in frequency chip-to-chip and chip-to-MegaZOR,
the entire system has a single resonance frequency with near-zero
currents at the spines, which are voltage antinodes. The frequency
mismatch causes elevated currents in the spines due to shift from ZOR
mode towards the first mode and an increase in total drive power. For
the worst case described above, our design was optimized to a
maximum 50\,mA variation of the current in the spines. The parasitic
current in the spine divides across multiple ribs of the on-chip
resonator. For the current design with about 100 ribs at the current
antinode carrying a nominal current of 5\,mA, the total resulting
variation of the bias current amplitude as seen by the circuit
junctions is less than 1\% and the input amplitude adjustment to the
system is only 1.5\,dB.

\begin{table}
\renewcommand{\arraystretch}{1.0}
\caption{Measured Margins for Individual Chips and for Chip Chains on
  the MCM with a Resonant Clock}
\label{428marg}
\centering
\begin{tabular}{c|l|l|l}
  {\bf Build} & {\bf Test Path} & {\bf Resonance} & {\bf AC Margins} \\
\hline
    & Chip A & 1.973\,GHz & 2.5\,dB \\
    & Chip B & 1.974\,GHz & 4.5\,dB \\
    & Chip C & 1.982\,GHz & 5.0\,dB \\
  1 & Chip D & 1.976\,GHz & 3.0\,dB \\
    & MCM Chips A-B & & untestable$^*$\\
    & MCM Chips C-D & & 4.8\,dB \\
    & MCM Chips A-B-C & & untestable$^*$ \\
\hline
    & Chip A & 2.019\,GHz & 4.0\,dB \\
    & Chip B & 2.011\,GHz & 4.0\,dB \\
    & Chip C & 2.019\,GHz & 3.5\,dB \\
  2 & Chip D & 2.031\,GHz & 4.5\,dB \\
    & MCM Chips A-B & & 2.0\,dB \\
    & MCM Chips C-D & & 2.5\,dB \\
    & MCM Chips A-B-C & & 1.0\,dB \\
  \hline
  \multicolumn{4}{l}{$^*$Untestable due to signal continuity failures on MCM}
\end{tabular}
\end{table}

Two MCMs were built using the two sets of chips, entered in
Table~\ref{428marg}, with resonant frequencies within 0.45\% of each
other and within $\pm 1\%$ of the MegaZOR frequency of the
MCM. Measurement of the on-carrier PCM resonators showed good
targeting of the fabrication process, with speed-of-light within 1\%
of the design value.

Table~\ref{428marg} also lists the measured AC clock margins for the
pretested chips and for the synchronous data links between chips. Only
the short path between chips C and D was testable on Build 1 due to
signal continuity failures. Measured margins for this path are on par
with those of the individual chips, indicating no degradation due to
the MCM transition. The applied clock frequency was tuned to the mean
resonance of Chips C and D for this test. Build 2 shows functionality
of all links between the four chips. The operating margin for the
links varied from 2.5\,dB for the short links to 1\,dB to the longest
link through Chips A, B, and C.

Simulation in ADS and HFSS, based on back-annotation of
the carrier physical layout, showed that the margin degradation is due
to inconsistent design of crossovers between I and Q carrier
resonators resulting in an imbalance between the four branches of
MegaZOR. As the imbalance between the two chips tested in Build 1 is
small, the effect on margins is minimal.  We conclude that the margin
degradation could be corrected in a redesign, and that synchronous
communication at the scale for four 10$\times$10\,mm$^2$ chips is
practical.

\section{Conclusion}
In this paper we have presented design and test results for
synchronous communication links between multiple chips on an MCM using
resonant clock distribution. These results advance state-of-the-art in
superconducting digital technology in multiple ways. It is the first
demonstration of superconducting data links on MCM involving multiple
chips. The system demonstrates the advantage of synchronous
communication in RQL technology overcoming the difficulties of clock
recovery and accumulated thermal jitter and parametric timing
uncertainty associated with RSFQ technology. We have shown that
digital data on MCM can be transmitted across 54\,mm between chips
with minimal hardware overhead as the LBW driver and receiver have
only six Josephson junctions, dissipating only 3\,fJ taking the
300\,W/W cooling overhead into account.  The energy transmitted per
bit is three orders of magnitude less than state-of-the-art CMOS
BoW-style data links for the same purpose and distance. The 2\,GHz
clock frequency in the current demonstration can be scaled up to
10\,GHz based on circuit simulation of the LBW driver, limited by
dispersion of the Nb interconnects. Extension to multi-bit bus
communication is straight forward, with total cross-sectional
bandwidth between chips limited only by bump pitch. All told, we have
enabled a critical interconnect functionality for future RQL systems
distributed on multiple chips using 3D-IC advanced packaging.

\begin{acknowledgments}

The authors acknowledge Andrew Brownfield for developing high
frequency test setup and assisting with circuit test, Paul Chang for
assisting with data reduction and analysis, and Vladimir Talanov for
valuable discussions and help with HFSS simulations. Authors also
acknowledge the contributions of the Lincoln Laboratory team,
particularly Rabindra Das and Sergey Tolpygo, in numerous technical
discussions regarding the MCM carrier and bonding process, and
assistance with design. The Northrop Grumman superconducting digital
EDA and PDK teams laid the foundation for MCM design and verification.
  
\end{acknowledgments}

\bibliography{mcm}

\end{document}